\documentclass[twocolumn,showpacs,preprintnumbers,amsmath,amssymb]{revtex4}

\usepackage{graphicx}
\usepackage{dcolumn}
\usepackage{bm}

\begin{document}

\title{MagnetoResistance of graphene-based spin valves.}

\author{L. Brey$^1$ and  H. A. Fertig$^{2,3}$}
\affiliation{1. Instituto de Ciencia de Materiales de Madrid
(CSIC),
Cantoblanco, 28049 Madrid, Spain\\
2. Department of Physics, Indiana University, Bloomington, IN 47405\\
3. Department of Physics, Technion, Haifa 32000, Israel}

\date{\today}

\begin{abstract}

We study the magnetoresistance of spin-valve devices using
graphene as a non-magnetic material to connect
ferromagnetic leads. As a preliminary step we first study the
conductivity of a graphene strip connected to metallic contacts
for a variety of lead parameters, and demonstrate that the
resulting conductivity is rather insensitive to them. We then
compute the conductivity of the spin-valve device in the parallel
and antiparallel spin polarization configurations, and find that
it depends only weakly on the relative spin orientations of the
leads, so that the magnetoresistance $MR$ of the system is very
small. The smallness of $MR$ is a consequence of the near
independence of the graphene conductivity from the electronic
details of the leads. Our results indicate that, although
graphene has properties that make it attractive
for spintronic devices, the performance of an
graphene-based spin-valve is likely to be poor.

\end{abstract}

\maketitle

\section{Introduction.}
Recently it has become possible to isolate an individual graphene
layer \cite{Novoselov_2004}, a two dimensional crystal of carbon
atoms packed in a honeycomb lattice. When deposited on top of a
doped dielectric substrate, the density of carriers moving in the
graphene sheet can be modified by applying a gate voltage
\cite{Novoselov_2005,Zhang_2005}. At low energies, carriers moving
in a graphene sheet obey the Dirac equation, so that graphene
offers the interesting possibility of studying the properties of
Dirac fermions. Apart from the interesting fundamental physics of
this new system, graphene is attracting attention as a promising
new material for microelectronic applications.

The conductivity of graphene tends to a minimum value when the
density of extra carriers tends to zero
\cite{Novoselov_2005,Zhang_2005,Tan_2007}. Theories predict a
universal value, $\sigma _u = 4 e^2 / \pi h$, for the conductivity
of a homogeneous and impurity-free undoped graphene sheet
\cite{Ludwig_1994,Ziegler_1998,Peres_2005,Katsnelson_2006,Stauber_2007}.
However experimental values of the minimum value of the conductivity
in graphene sheets is between five and ten times larger than the
theoretical  prediction. This discrepancy may occur because in
neutral graphene the system breaks up into electron and hole puddles
\cite{Martin_2007}, so that transport could occur through the
resulting hole-like and electron-like regions \cite{Cheianov_2007} .
The charge puddles may appear in order to screen impurities which
are invariably present \cite{Hwang_2007}, or may be induced by
ripples in the graphene sheets
\cite{Meyer_2007,Castro-Neto_2007,Juan_2007,Guinea_2007,Brey_2007b}.

In the ballistic regime, theoretical work \cite{Tworzydlo_2006,
Louis_2007,Schomerus_2006,Blanter_2006} has shown that the
conductivity of an intrinsic  graphene sheet of width $W$ and
length $L$ takes the universal value in the $W/L \rightarrow
\infty$ limit. Such a ballistic approximation is valid when the
mean free path of the carriers is larger than the sample
dimensions. This was the case, for example, for devices used in
Ref. \onlinecite{Miao_2007}, which confirmed that the conductivity
of wide and short graphene ribbons tends to the universal value $4
e^2 /\pi h$.

Graphene exhibits room temperature mobilities above $10^{5} cm ^2
/Vs$ \cite{Novoselov_2005}, implying that electrons in graphene
sheets can move very long distances without scattering. For short
range scatters the mean free path can be as large as 1000nm
\cite{Nomura_2007}. From these results we expect the ballistic
approximation to be appropriate to describe transport in graphene
nanoribbons  \cite{Chen_2007,Han_2007}. In addition the small
spin-orbit coupling of carbon atoms \cite{Huertas_2007} results in
a long spin lifetime for carriers in graphene sheets. This makes
graphene a very good candidate for microelectronic and spintronic
applications.

\begin{figure}
  \includegraphics[clip,width=9cm]{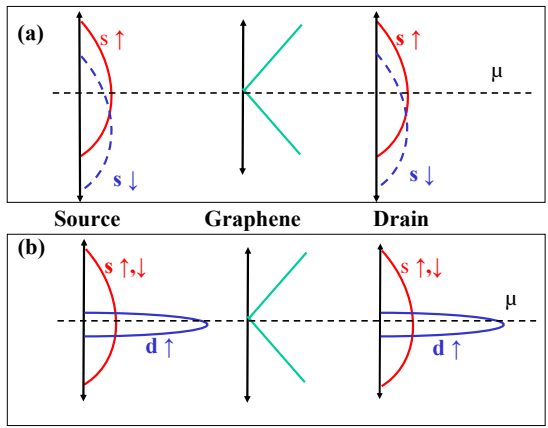}
  \caption{($Color$ $online$) Schematic representation of the spin valves studied in this work.
  We plot the density of states of the central graphene region and of the
ferromagnetic source and drain leads. In (a) the leads are
single orbital band metals where the center of the minority spin
band is shifted with respect to the majority spin band. (b)
corresponds to the case of a ferromagnetic transition metal  with
a spin polarized narrow $d$ band and a paramagnetic wide $s$ band.
The horizontal dashed line indicates the chemical potential $\mu$.
In this schematic picture $\mu =0$. In both cases, the spin
polarizations of the leads is illustrated in the parallel
configuration. For the antiparallel configuration the spin labels
of each band in the drain lead should be reversed.
  }
   \label{Figure1}
\end{figure}

In developing new spintronic devices, it is very important to find
non-magnetic  materials where a spin-polarized current can be
injected and flow without becoming depolarized. The most popular
existing spintronic devices are spin valves. These devices use the
fact \cite{Zutic_2004} that the electrical resistance of a
non-magnetic material connected to spin-polarized source and drain
leads depends strongly on their relative spin-orientation. Spin
valves are promising candidates for systems that may transform
spin information into electrical signals. These devices perform
best when the spin relaxation time of the non-magnetic material is
long, making graphene a good candidate for this component of a
spin-valve. Moreover, the combination of weak spin-orbit coupling
and low hyperfine interaction of the electron spin with the carbon
nucleus makes graphene a good candidate for other spintronic
applications such as spin qubits \cite{Trauzettel_2007},
three-terminal devices \cite{semenov_2007,haugen_2007}
and spin filters \cite{karpan_2007}.

Recently several groups have performed non-local four-probe
measurements \cite{Tombros_2007,Cho_2007,Ohishi_2007} of graphene,
connected to ferromagnetic electrodes, and have demonstrated the
presence of spin currents between injector and detector. These
experiments provide proof in principle of the possibility of
injecting a spin current into graphene, with a  spin relaxation
length larger than 2$\mu m$ at room temperature \cite{Tombros_2007}.
Moreover, a magnetoresistance of several hundred Ohms was observed
in a spin valve where graphene is contacted by two ferromagnetic
electrodes \cite{Hill_2006}. These experiments clearly demonstrate
the potential for graphene in spintronic devices.

In this work we study the magnetoresistance of a graphene-based spin
valve. This is a three component device, with a first ferromagnetic
lead used as a spin polarizer, a non-magnetic spacer -- graphene in
our case -- and a second ferromagnetic lead used as analyzer. We
consider two kinds of electrodes (see Fig. \ref{Figure1}). (a) A
single orbital band metal where the center of the minority spin band
is shifted with respect to the majority spin band in such a way that
the material is ferromagnetic. This is a simplified version of a
dilute magnetic semiconductor.  (b) A ferromagnetic transition metal
(e.g., cobalt) with a spin-polarized narrow $d$ band and a wide
paramagnetic $s$ band at the Fermi energy.

We assume the carrier mean free path in graphene is longer than the
dimension of the graphene part of the device, so that the transport
properties can be calculated in the \emph{ballistic approximation}.
In addition, since the spin-orbit coupling in graphene is very
small, it is appropriate to assume that the spin diffusion length is
sufficiently long that the carriers do not undergo spin flips while
traversing the graphene. Therefore we model the transport in terms
of two independent spin channels \cite{Fert_1976}. In this work we
will call this the independent current model.

This paper is organized as follows. In Section II we describe the
Green's function formalism for computing the conductance of a
ballistic system \cite{Datta_book}, and show how this leads to a
{\it conductivity} for undoped graphene.  In Section III we then
apply the formalism to compute the magnetoresistance of a wide
piece of graphene contacted by two single orbital band
ferromagnetic leads. In Section IV we discuss the case of
transition metal leads, modeled as conductors with two orbital
bands, one narrow and one wide.  We then compute the
magnetoresistance of this system. Finally, we summarize our
results in Section V.

\section{Conductance and Conductivity of a Graphene Strip}

We begin by first calculating the conductance of a graphene strip
connected to single orbital metallic leads [see Fig.
\ref{Figure1}(a)] for an arbitrary alignment of the lead band
bottoms and the Dirac points of the graphene system.

\subsection{System description.}
\begin{figure}
  \includegraphics[clip,width=9cm]{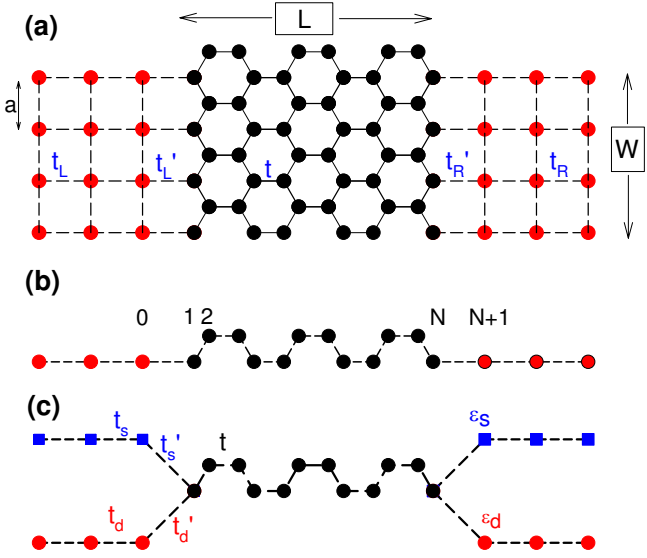}
  \caption{($Color$ $online$) (a) Structure of the spin valve studied in this
  work. A graphene sample is   connected to
  metallic leads  which are modeled as  commensurably matched square
  lattices. (b) Unit cell that it is periodically repeated in the
  $y$ direction in the case of a graphene strip connected to single
  orbital metallic leads. (c) Unit cell in the case of a graphene
  strip connected to ferromagnetic transition metal leads. Square
  (blue)
  points describe the $s$ atoms and  circle (red) points represent $d$
  atoms.
  }
   \label{Figure2}
\end{figure}

We consider a stripe of graphene as illustrated in Fig.
\ref{Figure2}(a), with lattice parameter $a$, attached to metallic
leads which are each modelled by a commensurably matched square
lattice with the same lattice parameter
\cite{Schomerus_2006,Blanter_2006}. The graphene strip has length
$L$ in the $x$ direction and is infinitely wide in the $y$
direction, i.e., $W \rightarrow \infty$. With this geometry, we
can define a unit cell, infinitely long in the $\hat{x}$
direction, that is periodically repeated in the $y$-direction
[Fig. \ref{Figure2}(b)]. In order to match the graphene sample to
the square lattice leads the number of carbon atoms in the unit
cell, $N$, should be a multiple of four, so that the length of the
graphene sample is $L=N/4\sqrt{3}a$.

\subsection{Hamiltonian}
The electronic properties of the graphene are described by a
tight-binding Hamiltonian with nearest neighbor hopping $t$ and
zero on-site energy. The left (L) and right (R) metallic leads are
also described by tight-binding Hamiltonians, with nearest
neighbor hopping amplitudes $t_L$ and $t_R$ and onsite energies
$\varepsilon _L$ and $\varepsilon_R$, respectively. The graphene
sample is connected to the left and right leads  with hopping
amplitudes $t_L ' $ and $t_R '$ respectively.

Because of the discrete translational invariance in the $y$
direction, we can label the eigenvalues and eigenvectors of the
Hamiltonian by a momentum $k_y$ in the $y$ direction that takes
values in the range $-\pi /a < k_y < \pi /a$. The limit $W
\rightarrow \infty$ is obtained by treating this as a continuous
variable. In order to simplify notation, in the rest of the paper
the wavevector $k_y$ will be given in units of $1/a$. With this
the electronic properties of the two dimensional problem is
reduced to a set of $k_y$-dependent one dimensional Hamiltonians
with on site energies (see Fig. \ref{Figure1})
\begin{eqnarray}
\epsilon _i & = &  \epsilon _L + 2 t _L \cos {k_y}\, \, \, \,\, \,
\, \, \, \, \, \, \, \, \, \,   \, \, \, \, \, \, \, \,
\textrm{if} \, \,\, \, \,\, \, \, \, \, \,  \, i \leq 0 \, \, \,
\, \, \, \,
 \nonumber \\
\epsilon _i & = &  \epsilon _R +2 t _R \cos {k_y}\, \, \, \, \, \,
\, \, \, \, \, \, \, \, \, \,   \, \, \, \, \, \, \, \,\textrm{if}
\, \, \,  \, \, \, \, \, \, \,
i \geq N+1 \nonumber \\
\epsilon _i & = &0 \, \, \ \, \, \, \, \, \, \, \, \, \, \,\, \, \
\, \, \, \, \, \, \, \, \, \, \,\, \, \ \, \, \, \, \, \, \, \, \,
\, \, \, \, \ \, \, \, \, \, \, \, \, \, \, \,  \textrm{if} \, \,
\, \, \, \, \, \, \,  1< i < N, \label{onsite}
\end{eqnarray}
and hopping amplitudes $t_i$ between sites $i$ and $i+1$
\begin{eqnarray}
t_i &=& t _L \, \, \, \, \textrm{if} \, \, \, i < 0 \, \, \
\nonumber \\
t_i &=& t _L '  \, \, \, \, \textrm{if} \, \, \, i =0  \, \,
\ \nonumber \\
t_i &=& t + t e ^{i k_y} \, \, \, \, \, \, \textrm{if} \, \, \,
i=1,5,9... \, \, \, \, \, \, \, \, \,  \, \, \,\textrm{and} \, \,
\, \, 0 < i < N-1 \nonumber \\  &=& t  \, \, \, \, \, \, \,\, \,
\,\,\, \, \, \, \, \,\, \, \, \, \, \,\, \,\textrm{if} \, \, \,
i=2,6,10... \, \, \, \, \, \, \, \, \, \textrm{and}  \, \, \, 0 <
i < N-1 \nonumber \\&=& \,t + t e ^{-i k_y} \, \,  \textrm{if}
\,\,\, i=3,7,11... \, \, \, \, \, \, \, \, \, \, \textrm{and}  \,
\, \, 0 < i < N-1 \nonumber \\&=& t  \, \, \, \, \, \, \, \, \, \,
\, \,\, \, \, \, \, \, \, \, \, \,\, \, \, \, \textrm{if}   \,\,\,
i=4,8,12... \, \, \, \, \, \,  \, \, \, \, \, \textrm{and}  \, \,
\, 0 < i < N-1 \nonumber \\ t_i &=& t _R ' \, \, \, \, \textrm{if}
\, \, \, i = N\, \, \
\nonumber \\
t_i &=& t _R  \, \, \, \, \textrm{if} \, \, \, i >N  \, \,.
\label{tild_t}
\end{eqnarray}
\subsection{Conductance}
The conductance per spin channel of the system takes the form
\begin{equation}
G(\mu)= \sum_{k_y} g ^{1D} (\mu, 2t _L \cos k_y+ \epsilon _L)
\label{conductance}
\end{equation}
where $\mu$ is the chemical potential and $g ^{1D} (E,\epsilon _0)$
is the conductance of the one dimensional chain with an on-site
energy $\epsilon _0$ in the left electrode, evaluated for Fermi
energy $E$.  (Note that $g ^{1D}$ implicitly depends on the on-site
energy in the right lead as well.)
The one dimensional conductance is given by \cite{Datta_book}
\begin{equation}
g ^{1D} (\mu, 2t _L  \cos k_y+ \epsilon _L)= \frac {e^2}{h} \,
\Gamma _1 (\mu) \, \Gamma _N (\mu) |G_{1,N}(\mu)|^2 \label{1dcond}
\end{equation} where $\Gamma _1 (E) = i
( \Sigma _1 -\Sigma _1 ^+)$ and $\Gamma _N (E) = i ( \Sigma _N
-\Sigma _N ^+)$ are the injection ratios, and  $\Sigma _1$ and
$\Sigma _N$ are the self-energy terms due to the leads attached at
sites 1 and $N$ respectively \cite{Datta_book},
\begin{equation}
\Sigma _{1 (N)} (\mu)= t_{L(R)}' \, \,  e^{-ik_{L(R)}}
\label{self_chain}
\end{equation}
with
\begin{equation}
k_{L(R)}=\arccos{\left (\frac {\mu-\epsilon_{L(R)}-2 \,
t_{L(R)}\cos{k_y}}{2 \, t _{L(R)}} \right )} \, \, \, .
\label{klkr}
\end{equation}
In Eq.\ref{1dcond} $G_{1,N} (\mu)$  is the effective Green's
function for the system, evaluated in the graphene region:
\begin{equation}
G_{1,N} (\mu) = \frac {1}{ \mu-H_c-\Sigma _1 (\mu) -\Sigma _N
(\mu)}. \label{G1N}
\end{equation}
Here $H_c$ is the one dimensional $k_y$-dependent Hamiltonian of
the graphene sample described by the on-site energies and hopping
amplitudes given  in Eqs. \ref{onsite} and \ref{tild_t}.

Finally, the conductivity of the device is related to the
conductance through geometrical factors,
\begin{equation}
\sigma (\mu) = \frac L W G (\mu) \, \, \,.
\end{equation}
In two dimensions, conductivity and conductance have the same
units. However, the former is only useful in systems where its
value is sample-size independent for large $L$ and $W$.  Usually
this is only the case for diffusive systems. Remarkably, as
has been pointed out previously
\cite{Tworzydlo_2006,Schomerus_2006} and as we demonstrate
explicitly below, for undoped graphene it is true in the {\it ballistic}
limit, provided we take $W \rightarrow \infty$ before taking $L
\rightarrow \infty$.

\begin{figure}
  \includegraphics[clip,width=9cm]{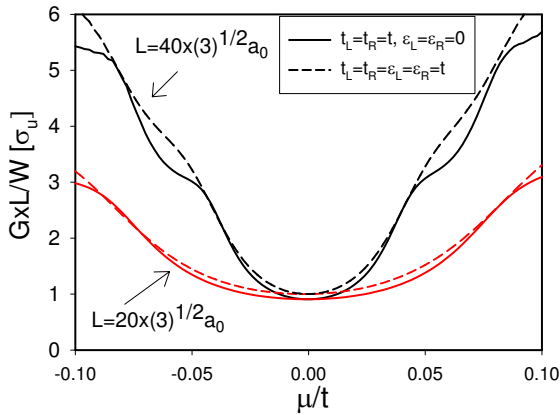}
  \caption{($Color$ $online$) Conductivity in units of $\sigma _u = 4/\pi (e^2/h)$, as
  a function of the chemical
  potential for two different graphene sample lengths and two sets of
  parameters describing the lead band structures.  The conductivity
  shows a minimum at the Dirac point which is independent of the
  sample length.
  }
   \label{Figure3}
\end{figure}

In Fig. \ref{Figure3} we plot $G(\mu) \! \times \!  L /W$  in
units of the universal conductivity for two sets of parameters for
the leads, in the limit of large $W$, and for two different values
of $L$.  One may observe from these results that except for
$\mu=0$ the conductance is essentially independent of $L$.
Interestingly, the conductance is relatively insensitive to the
details of the metallic leads; its overall shape is mostly
determined by the graphene density of states.  In particular the
roughly linear rise as $\mu$ moves away from zero may be
understood as reflecting the linear density of states of graphene.
The shoulders in the conductance riding on this linear background
correspond to Fabry-Perot resonances due to the finite length of
the graphene sample \cite{MuñozRojas_2007}. The amplitude of these
resonances is determined by the parameters of the leads.

In the case of intrinsic (i.e., undoped) graphene -- $\mu$=0 --
the electronic transport behaves as if it is diffusive, and the
conductivity is well-defined. For $t_L=t_R=t_L'=t_R'=t$ and
$\varepsilon _L = \varepsilon _R =0$ we obtain $\sigma =
\sqrt{3}/2 (e^2/h)$, in agreement with Ref.
\onlinecite{Schomerus_2006}.  This value changes when the centers
of the energy bands in the leads change.  The conductivity reaches
the universal value, $\sigma _u =4 / \pi (e^2 /h)$, when
$\varepsilon _L =t _L$ and $\varepsilon _R =t_R$
\cite{Tworzydlo_2006,Schomerus_2006}.

\begin{figure}
  \includegraphics[clip,width=9cm]{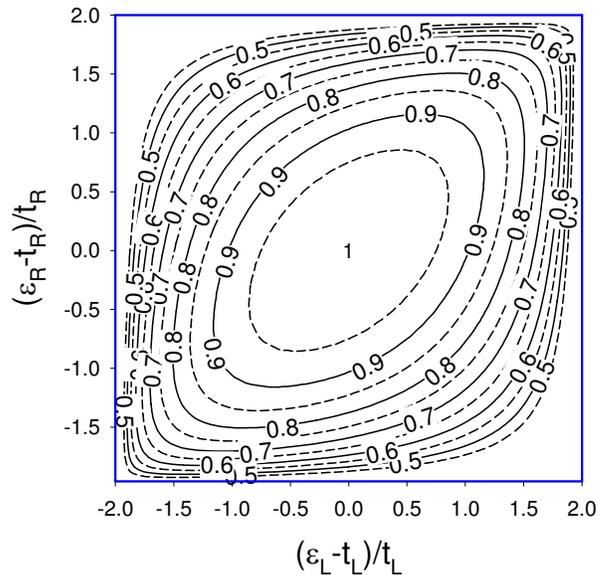}
  \caption{($Color$ $online$) Conductivity evaluated at the Dirac point, $\mu=0$,
  in units of $\sigma _u = 4 /\pi (e^2/h)$, as a function of the
  source and drain leads parameters.
  }
   \label{Figure4}
\end{figure}

The $\mu=0$ case may be worked essentially analytically, because
transport is dominated by values of $k_y$ in the vicinity of the
Dirac points, $(0,\pm \frac {4 \pi }{3 a})$. In this case and for
large values of $L$ it is possible to explicitly evaluate the
transmission probability associated with $H_C$ for each valley and
spin channel using a transfer matrix method, with the result
\begin{equation}
T_{k_y} (\mu\! =\! 0) \! = \! \frac {4 \xi t^2 \sin{k_L}
\sin{k_R}}{ \xi ^2 e ^{2 L k_y} \!+ \!  t ^4 e ^{-2 L k_y}  \! -
\! 2 t^2 \xi \cos{(k_L \! \! + \! \! k_R)}}. \label{trans_amp}
\end{equation}
In this expression the parameter  $ \xi = (t_L ^{'2} t_R
^{'2})/(t_L t _R) $ contains the information about the coupling to
the leads, the wavevector $k_y$ is defined with respect the Dirac
point,  and $k_L$ and $k_R$ are defined in Eq. \ref{klkr}. For
$\mu$ at the Dirac points there are {\it no} individual
propagating modes in the graphene sample connecting the metallic
leads at any finite value of $k_y$, and the transmission
probability vanishes in the limit $L \rightarrow \infty$.

By integrating the transmission amplitude (Eq.\ref{trans_amp})
over $k_y$ we obtain the $\mu=0$ conductance,
\begin{equation}
G \! = \frac {e^2}{h}  \frac {2 g_s g_v} {\pi} \frac {W}{L}
\frac{\sin {k_L} \sin{k_R}}{\sin{(k_L \! \! + \! \!k_R)}} \arctan
\left ( \frac {\sin{(k_L\! \! + \! \! k_R)}}{1 \! \! - \! \!
\cos{(k_L \! \! + \! \! k_R)}} \right ) \label{cond_ana}
\end{equation}
where $g_s$=2 and $g_v$=2 are the spin and valley degeneracies
respectively. Remarkably, although all the modes are evanescent,
the total conductance falls off only as $1/L$, so that the
electrical transport is Ohmic.  This behavior arises because the
effective length scale of the evanescent modes are each $1/k_y$,
which is arbitrarily large as $k_y \rightarrow 0$, resulting in
the relatively weak $1/L$ behavior for $G$. (Alternatively, the
finite value of the transmission probability near the Dirac points
may be understood in terms of virtual electron-hole pair
excitations near zero energy in the graphene region
\cite{Tworzydlo_2006}.) By contrast, for systems in which $\mu$
may lie in a gap, the effective length scale for wavefunctions is
bounded by a distance that is determined by the difference between
$\mu$ and the bottom of the conduction band for the ``conducting"
region.  Because of this maximum length scale $G$ falls off
exponentially with $L$, and the conductivity as well as the
conductance vanish as $L \rightarrow \infty$.  Thus clean graphene
is rather unique in displaying Ohmic behavior.

Another remarkable feature of this result is that the conductivity
does not depend on the couplings $t_L' $ and $t_R'$ of the
metallic leads with the graphene sample. This result is only
possible in the limit $W \rightarrow \infty$, since for any finite
value of $W$ the two leads become fully disconnected if either of
these parameters vanishes, and the conductance must vanish.
However, in the infinite width limit where the momentum sum
becomes an integral, the parameter $\xi$ through which $t_L' $ and
$t_R'$ enter scales out, and the final result is independent of
these parameters.

Finally, it is interesting to examine the dependence of the
conductivity on the parameters specifying the leads.  From the
definition of $k_L$ and $k_R$ one can see that the conductance
depends on the lead parameters only through the combinations
$(\epsilon _R - t_R)/t_R$ and $(\epsilon _L - t_L)/t_L$. In Figure
\ref{Figure4} we plot the conductivity, $\sigma = G \times  L /W$,
as a function of these combinations. The conductivity is maximized
when $t_L=\varepsilon _L$ and $t_R=\varepsilon _R$, independent of
the values of $t_L$ and $t_R$. Because of conservation of energy
and of the transverse momentum, $k_y$, transport through the
graphene ribbon when $\mu=0$ is possible only when the band center
of the left and right leads are in the intervals $2 t_L<|\epsilon
_L - t_L|$ and $2 t_L<|\epsilon _L - t_L|$ respectively. In Ref.
\onlinecite{Schomerus_2006} Schomerus suggests that the
conductivity of a graphene sample connected to metallic leads is
maximum when the self-energies of electrons in the leads have the
same value as the self-energy of bulk graphene with energy at the
Dirac point ($-i t$). In the square lattice the self-energies of
the incoming and outcoming electrons, evaluated at the momentum
and energy of the Dirac point, are $-t_L e^{ik_L}$ and $-t_R
e^{ik_R}$, respectively. Our results indicate that when the {\it
real} part of the lead self-energies $\Sigma_1(\mu=0)$ and
$\Sigma_N(\mu=0)$ (Eq. \ref{self_chain}) vanish, the conductivity
of the system is maximized. This condition is similar to but less
restrictive than that proposed in Ref.
\onlinecite{Schomerus_2006}, and is related to the independence of
the graphene universal conductivity from the hopping amplitude
$t$.

\section{Magnetoresistance: Single Band Leads}
In this section we apply the results derived above to find the
magnetoresistance of a spin valve device with a graphene strip at
its center. In  a single band ferromagnetic metal, the centers of
the spin up ($\varepsilon _{0 ,{\uparrow}}$) and spin down
($\varepsilon _{0 ,{\downarrow}}$) bands are shifted, as
illustrated in Fig. \ref{Figure1}(a). This implies a relative spin
polarization $P$ of the carriers at the Fermi energy $\mu$ given
by
\begin{equation}
P \!  = \! \frac { t_{\downarrow} \mathbb{K} \left (\sqrt{1 \! -
\! (\frac { \mu \! - \! \varepsilon _{0,\uparrow}}{2t_{\uparrow}}
) ^2 }\right )- t_{\uparrow} \mathbb{K} \left (\sqrt{1 \! - \!
(\frac { \mu \! - \! \varepsilon _{0,\downarrow}}{2t_{\downarrow}}
) ^2 }\right )} {t_{\downarrow}\mathbb{K} \left (\sqrt{1 \! - \!
(\frac { \mu \! - \! \varepsilon _{0,\uparrow}}{2t_{\uparrow}} )
^2 }\right )+ t_{\uparrow}\mathbb{K} \left (\sqrt{1 \! - \! (\frac
{ \mu \! - \! \varepsilon _{0,\downarrow}}{2t_{\downarrow}} ) ^2
}\right )}, \label{polarization}
\end{equation}
where $\mathbb{K} $ is the complete elliptic integral of the first
kind, and $t_{\uparrow}$ and $t_{\downarrow}$ are the hopping
matrix elements in the spin up and spin down channels
respectively. Eq. \ref{polarization} is obtained directly from the
density of states at energy $E$ of a square lattice with hopping
parameter $t$ and on-site energy $\varepsilon _0$, which is given
by \cite{e83}
\begin{equation}
\rho(E)=\frac {1}{ \pi ^2 t} \Theta (2 t -|E-\varepsilon _0|)
\mathbb{K} \left (   \sqrt{1-\frac {(E-\varepsilon _0)^2 }{4
t^2}}\right ).
\end{equation}

The transport through a non-magnetic material connected to
ferromagnetic metals is expected to depend strongly on the
magnitude and relative orientation of the polarizations of the
leads. The magnetoresistance ($MR$) is defined as the relative
change of the resistance when the relative spin  orientation of
the leads changes from parallel, $R_{para}$, to antiparallel,
$R_{anti}$. In the ballistic approximation the MR can be written
as
\begin{equation}
MR= \frac {R_{anti}-R_{para}}{R_{anti}}=\frac {G_{para}-G_{anti}}
{G_{para}} \label{MR} \, \, \,
\end{equation}
where $G_{para}$ and $G_{anti}$ are the conductances of the system
when the relative polarization of the leads is parallel or
antiparallel respectively.

When the non-magnetic material is an insulator the transport is
through tunneling processes. Assuming that tunneling transport is
proportional to the the density of states at the Fermi energy,
Julliere \cite{Julliere} proposed the following expression for the
tunneling MR:
\begin{equation}
TMR= \frac{2 P^2}{1+P^2}. \label{Julliere}
\end{equation}
Julliere's expression works rather well for tunneling MR even for
materials with complicated band structures \cite{Brey_2004b}.  It
predicts correctly that  the MR increases as the leads become more
spin polarized. In this section we study  the MR when the
non-magnetic material is graphene, and analyze the results as a
function of the spin polarization of the leads.

In the independent current model, and using the ballistic
approximation, $G_{para}$ is the sum of the spin up conductance,
Eq. \ref{conductance} evaluated with $\varepsilon _L = \varepsilon
_R =\varepsilon _{0,\uparrow}$, and the spin down conductance, Eq.
\ref{conductance} evaluated with $\varepsilon _L = \varepsilon _R
=\varepsilon _{0,\downarrow}$. In the case of antiparallel spin
polarization of the leads $G_{anti} $ is the sum of Eq.
\ref{conductance} evaluated with $\varepsilon _L =\varepsilon
_{0,\uparrow}$ and $\varepsilon _R =\varepsilon _{0,\downarrow}$
and Eq. \ref{conductance} evaluated with $\varepsilon _L
=\varepsilon _{0,\downarrow}$ and $\varepsilon _R =\varepsilon
_{0,\uparrow}$.

\begin{figure}
  \includegraphics[clip,width=9cm]{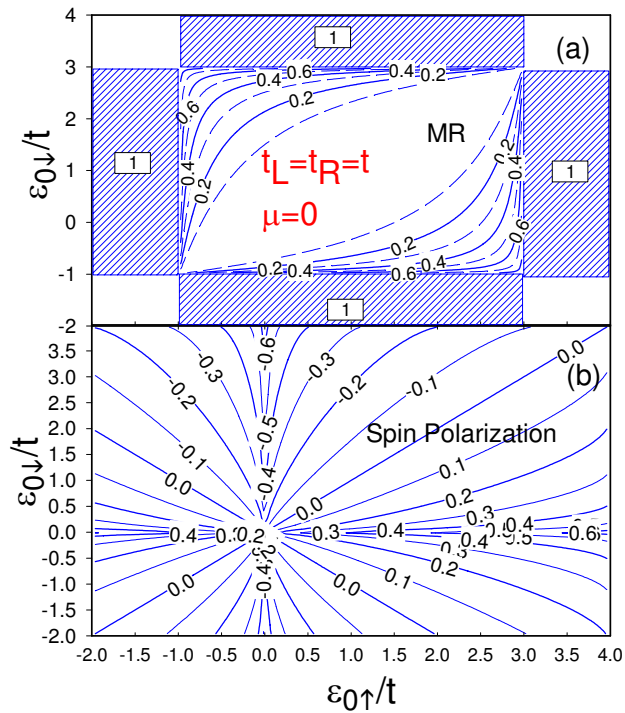}
  \caption{($Color$ $online$) (a) Magnetoresistance ($MR$) and (b)  spin
  polarization at the Fermi energy, as a function  of  the centers of the  spin
  bands, of a graphene spin valve. The device consists
  of  a long
  graphene slab separating single  band ferromagnetic
  metals, Fig.\ref{Figure1}(a). The chemical potential
  $\mu$=0, and  $t_L=t_R=t$.
  }
   \label{Figure5}
\end{figure}

In Fig. \ref{Figure5} we plot the magnetoresistance $MR$ of a
graphene based spin valve as function of the positions of the
center of the spin up and spin down bands. We consider only
intrinsic graphene ($\mu =0$), and the hopping parameter in the
metallic leads is taken equal to that in the graphene part of the
device, $t_L$=$t_R$=$t$. The graphene slab is extrapolated to
infinite length, although for $L$ larger than about $20 a$ the
results are essentially the same.

We observe that in general  the magnetoresistance is  small. Only
when the parameters are such that one of the spin bands is close
to or in the forbidden transport region, $2t<|\varepsilon
_{0\sigma}-t|$ does one find $MR$ to be a significant fraction of
one, and when one of the spin bands is in the forbidden region
then $MR$ of course reaches its maximum possible value. Thus in
order to get a moderate value of the magnetoresistance a large
shift between the center of the spin bands in the leads is needed.

The reason for the smallness of $MR$  is the weak dependence of the
graphene conductance on the parameters of the leads [see Figs.
\ref{Figure3} and \ref{Figure4}.] This weak dependence,
particularly on the density of states of the incoming electrons,
also implies an absence of any strong correlation between the spin
polarization of the incoming electrons [Fig. \ref{Figure5}(b)] and
the magnetoresistance [Fig. \ref{Figure5}(a)]. By comparing the
values of the polarization and the magnetoresistance we observe
that the Julliere expression, Eq. \ref{Julliere}, fails to
describe the magnetotransport properties of graphene-based
transistors.

Note also that because the conductivity of undoped graphene is
independent of the tunneling amplitude connecting the graphene with
the ferromagnetic metallic leads (Eq. \ref{cond_ana}), the
magnetoresistance  is also unaffected by changes in the quality of
the connection between the leads and the graphene. This indicates
that the smallness of the magnetoresistance in graphene-based
devices is not due to a conductivity mismatch, as is the case in
metal-semiconductor diffusive junctions \cite{Schmidt_2000}, but
rather is due to the \emph{universal} minimum conductivity of
graphene, so that a ``shutoff'' of one spin channel is difficult to
achieve.

\begin{figure}
  \includegraphics[clip,width=9cm]{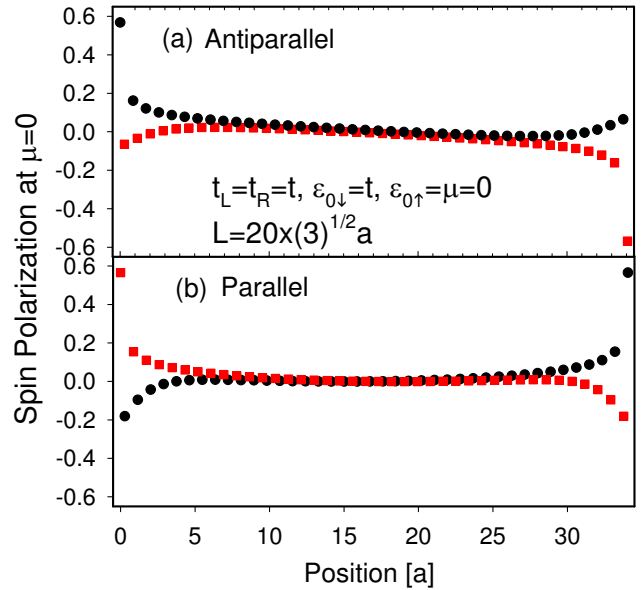}
  \caption{($Color$ $online$)
  Spin polarization, evaluated at the Fermi energy, as a function of the position
  along the graphene slab of a graphene-based spin valve. We plot the $\mu=0$
  case. (a) and (b) correspond to  antiparallel and parallel configurations respectively.
  Square and dot symbols indicate the spin polarization on the two types of lattice
  sites of the
  honeycomb lattice.
  The
  band structure parameters are indicated in the inset of the
  figure. The length of the graphene slab is $L=20\times \sqrt {3}a$.
  }
   \label{Figure6}
\end{figure}

In Fig. \ref{Figure6} we have plotted the spin polarization as a
function of position in our graphene-based spin valve. For this
calculation the lead parameters were chosen so that the spin
polarization in the leads was approximately 60$\%$. Several points
are worth noting. (1) The spin polarization decays to a zero over a
length scale of about 20 lattice parameters. (2) The spin
polarization  is oriented in {\it opposite directions} for electrons
on different sublattices. (3) The total induced spin in the graphene
region vanishes. These effects are explained  by the strong tendency
for local magnetic moments in graphene to orient ferromagnetically
for sites on the same sublattice, and antiferromagnetically for
sites on different sublattices \cite{Brey_2007}. We note that this
result does depend on the fact that in our chosen geometry, zigzag
edges present themselves to the leads, so that all the graphene
sites contacting the leads are on the same sublattice.  The
sensitivity of the electronic states to the edge geometry is a
well-known property of graphene
\cite{nakada_1996,Brey_2006,brey_fertig_2006}

\begin{figure}
  \includegraphics[clip,width=9cm]{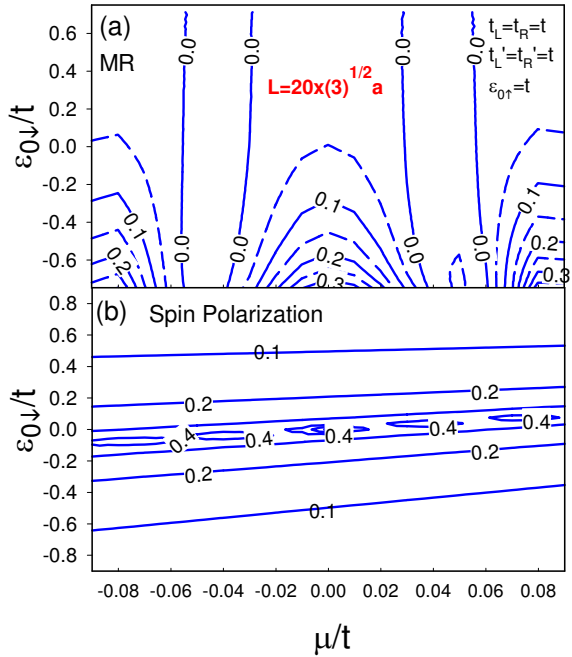}
  \caption{($Color$ $online$)
(a) Magnetoresistance ($MR$) and (b)spin polarization at the Fermi
energy as function of the chemical potential in the graphene layer
and the spin down band center. The center of the spin up band is
located in the optimal conductivity  position, $\varepsilon
_{0,\uparrow}$=$t$. We take $t_L$=$t_R$=$t'_L$=$t'_R$. The
length of the graphene slab is set to $L=20\times \sqrt {3}a$.  For
larger values of $L$ the
results are essentially identical.
  }
   \label{Figure7}
\end{figure}

Finally, we analyze the magnetoresistance for {\it doped} graphene
in the three stripe geometry. In this case the transport through the
graphene part of the device is ballistic in the usual sense, and the
$MR$ has to be calculated as the relative change in the conductance,
Eq. \ref{MR}. In Fig. \ref{Figure7}(a) we plot the $MR$ for this
case, as a function of the chemical potential in the graphene, and
of the center of the spin down band. In addition we plot the spin
polarization at the leads. As in the case of undoped graphene the MR
is small (although not quite so small as in the undoped case), and
only moderate values of $MR$ result from large shifts between the
centers of the spin up and spin down bands. From comparing the two
panels of Fig. \ref{Figure7}, we again observe that the values of
$MR$ and the spin polarization at the leads are largely
uncorrelated.

\section{Magnetoresistance: Ferromagnetic Transistion Metal Leads}
Energy bands for ferromagnetic transition metals can be described
approximately by using two bands, a $d$-band characterized by a
width $4t_d$ and center position $\varepsilon _d$, and an $s$-band
with width $4t_s$ and center at $\varepsilon _s$
\cite{Harrison_book}. Because the shape of the atomic orbitals,
the $s$-band is much wider that the $d$-band. Cobalt, for example,
is a transition metal with a minority spin $d$-band shifted up in
energy relative to a majority spin $d$-band. The $s$-band is
nearly spin-unpolarized, and the Fermi energy crosses both the
majority spin $d$-band and the unpolarized $s$-band, as
illustrated in Fig.\ref{Figure1}(b). For studying the transport
properties of such systems, it is a good approximation to neglect
the minority spin $d$-band located at high energy.

We wish to consider the same device as studied in Section III with
leads characterized by bandstructures of this sort.  This can be
modeled by assigning two orbitals to each site in the leads, one
for the $s$-band and the other for the $d$-band.  We can then
assign on-site energies $\varepsilon_{s,d}$ respectively for the
two orbitals, and hopping matrix elements $t_s$ and $t_d$.  These
orbitals then connect to the $p_z$ carbon graphene orbitals
through hopping amplitudes $t_s '$ and $t_d '$ respectively.
Although cobalt is a three dimensional metal, for simplicity we
treat it here as two dimensional.  The resulting one-dimensional
problem that one needs to treat for each $k_y$ is illustrated in
Fig. \ref{Figure2}(c). Finally, for this section we will restrict
our discussion to the case of intrinsic graphene, $\mu=0$.

\subsection{Conductivities}

As there are two band orbitals in the transition metal leads, in
order to compute the conductivity with different spin polarization
orientations in the leads we will need to evaluate several partial
conductivities, representing transmission between  different
combinations of orbitals.  These are:
\begin{itemize}
\item The conductivity across a graphene strip attached to single
band spin-polarized metals, with the bands in the two leads
centered at the same energies and having equal bandwidths,
\begin{equation}
\sigma _{\nu,\nu} = \frac L W  \frac 1 {g_s} G (k _{\nu}),
\label{sss}
\end{equation}
where $G(k _{\nu})$ is the conductance (Eq. \ref{cond_ana})
evaluated with $k_L$=$k_R$=$k_{\nu}=\arccos(\frac
{t_{\nu}-\varepsilon_{\nu}}{2t_{\nu}})$. This contribution enters
when we consider the transmission of spin-minority electrons in
the parallel configuration, so that the $d$-bands of the minority
spin are high in energy in both leads and are irrelevant.  Thus
only the band index $\nu=s$ for this contribution will be relevant
in our calculation. \item The conductivity of the device where the
source and drain metals consist of two conducting bands, $s$ and
$d$. In this case the conductivity per spin channel is
\begin{eqnarray}
\sigma _{sd,sd} & = & \frac {e^2} h \frac {4 g_v} {\pi} \left (
t_s \sin k_s + t_d \sin k_d \right )^2  \nonumber \times \\ &
\times & \frac { \arctan \left ( \frac
{\sqrt{4|a|^4-(a^2+a{^{*2}})^2}}{2 |a|^2-a^2 - a^{*2}} \right )}
{\sqrt{4 |a|^4-(a^2+a^{*2})^2}} \label{ssdsd}
\end{eqnarray}
with
\begin{equation}
a=t_s e ^{i k_s}+t_d e ^{ik_d}.
\end{equation}
This contribution enters for spin-majority electrons when the lead
polarizations are parallel. \item Finally it is necessary to
evaluate the conductivity in the case where the only available
band in the source metal  is  the $s$-band, whereas both bands,
$s$ and $d$, are available in the drain lead, or vice-versa. These
two situations arise when the lead polarizations are antiparallel.
The required conductivities are both
\begin{eqnarray}
\sigma _{s,sd} = \sigma _{sd,s}& = & \frac {e^2} h \frac {4 g_v}
{\pi} \left ( t_s
^2  \sin ^2  k_s + t_s  t_d \sin k_d  \sin k_s \right )  \nonumber \times \\
& \times & \frac { \arctan \left ( \frac
{\sqrt{4|b|^4-(b^2+b^{*2})^2}}{2 |b|^2-b^2 - b^{*2}} \right )}
{\sqrt{4 |b|^4-(b^2+b^{*2})^2}} \, \, \, ,  \label{sssd}
\end{eqnarray}
with
\begin{equation}
b=(t_s e ^{i k_s}+t_d e ^{ik_d}) e ^{i k_s} t _s .
\end{equation}
\end{itemize}

The above conductivities are independent of the values of the
contact hopping  amplitudes  between the graphene orbitals and the
$s$ and $d$ orbitals of the transition metal leads.

\begin{figure}
  \includegraphics[clip,width=9cm]{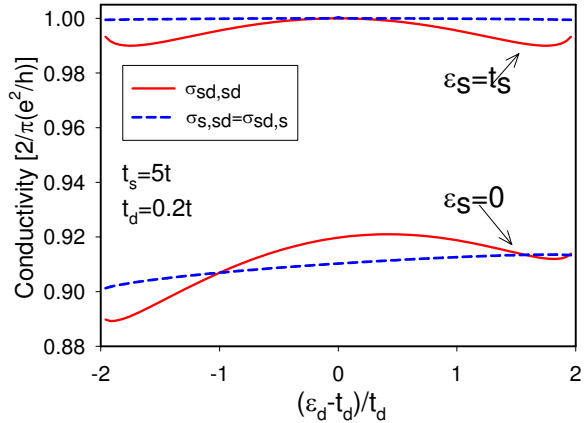}
  \caption{($Color$ $online$) Conductivity per spin channel of a
  a graphene slab connected to metallic leads with
  two ($s$ and $d$) bands each, $\sigma _{sd,sd}$ and
  to a two orbital ($s$ and $d$) source lead and a single orbital ($s$) drain lead,
  $\sigma_{sd,s}$. We plot the conductivity as a function of the
  band center of the $d$-band. We assume that the $s$-band is much
  wider than the $d$ band, taking $t_s$=5$t$ and $t_d$=0.2$t$. Plotted are the
  $\varepsilon _ {\nu}$=$t_{\nu}$  and the   $\varepsilon _ {\nu}$=0
  cases. Note that in all the cases the conductivity depends weakly
  on the parameters of the metallic leads.
  }
   \label{Figure8}
\end{figure}

A very interesting result which emerges from Eqs. (\ref{sss}),
(\ref{ssdsd}) and (\ref{sssd}) is that the conductivity of intrinsic
graphene is nearly independent of the number of bands in the drain
and source leads. For example, if we take $\varepsilon _s$=$t_s$ and
$\varepsilon _d$=$t_d$ we find that all the conductivities are the
same: $\sigma _{ss}$=$\sigma_{dd}$=$\sigma_{sd,s}$=$\sigma_{sd,sd}=
2/\pi \frac {e^2}{h}$. (Note that we are evaluating the conductivity
per spin channel). Similarly, if we center all the bands at zero
energy, $\varepsilon _s$=$\varepsilon _d$=0, all  the conductivities
take the value $\sqrt {3}/4 \frac {e^2}{h}$, independent of the
hopping parameters.  Fig. \ref{Figure8} illustrates this behavior
over a range of lead parameters. These results reflect the
remarkably weak dependence of the graphene conductivity on the
electronic structure of the contacts, and in particular on the
density of states of the metallic leads at the Fermi energy.

\subsection{Magnetoresistance}

For computing the magnetoresistance we have to evaluate the
conductivity in the parallel and  antiparallel spin polarization
configurations  of the leads.

In the parallel configuration, majority spin electrons in the $s$
and $d$ bands of the metallic source are injected   in the
graphene slab and received in the majority spin  $s$ and $d$ bands
of the metallic drain. The minority spin electrons can go just
from the source $s$-band to the drain $s$-band. Therefore the
conductivity is
\begin{equation}
\sigma_{para} = \sigma _{sd,sd}+\sigma_{s,s} \label{s_para}.
\end{equation}
In the antiparallel configuration the majority spin electrons in
the source can be in the $s$ or $d$ band.  Upon passing through to
the drain lead these are now minority electrons, which can only
reside in the $s$-band.  The inverse situation occurs for the
minority spin carriers in the source lead.  Thus the conductivity
in the antiparallel configuration takes the form
\begin{equation}
\sigma_{anti} = \sigma _{s,sd}+\sigma_{sd,s} \, \, \,
.\label{s_anti}
\end{equation}
Since by symmetry $\sigma _{s,sd}$=$\sigma_{sd,s}$, the
magnetoresistance  takes the form
\begin{equation}
MR=\frac {\sigma _{sd,sd}+\sigma_{s,s}-2\sigma _{s,sd}}{\sigma
_{sd,sd}+\sigma_{s,s}}.
\end{equation}

\begin{figure}
  \includegraphics[clip,width=9cm]{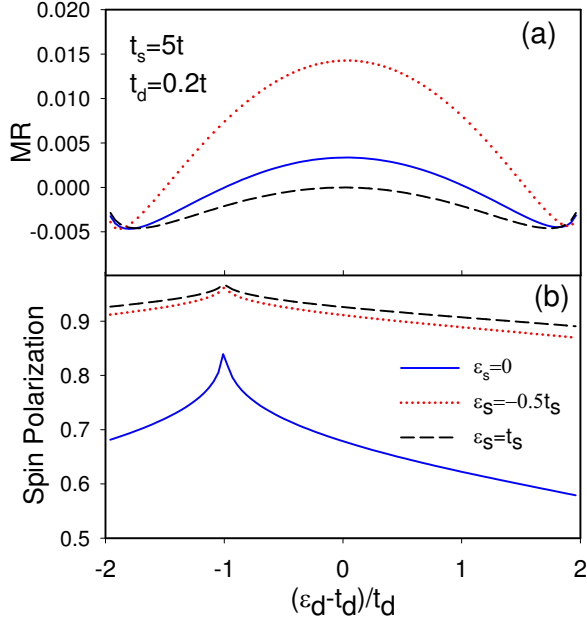}
  \caption{($Color$ $online$) (a) Magnetoresistance of a graphene strip connected
   to ferromagnetic transition metal leads.
  (b) Spin polarization at the Fermi energy of the ferromagnetic leads.
  The $d$ band is assumed to be much wider than the $s$ band,
  $t_s$=$5t$ and $t_d$=$0.2t$, with $t$ the graphene hopping
  parameter. We present the results for different positions of the
  center of the $s$ band.}
   \label{Figure9}
\end{figure}

In Figure \ref{Figure9}(a) we plot our calculated magnetoresistance
for this system. We assume the $s$-band to be much wider than the
$d$-band, and plot the results as functions of the energy center of
the $d$-band, and for different positions of the center of the
$s$-band. In Figure \ref{Figure9}(b) we plot the corresponding spin
polarization of the ferromagnetic leads.  From Fig. \ref{Figure9} we
immediately see that the magnetoresistance is generically very
small, which we can again understand as a consequence of the
relative insensitivity of the conductivity to the details of the
leads.  Interestingly, this same insensitivity means that there is no
simple relationship between the value of the spin polarization in
the leads and $MR$, as can be seen by comparing Figs.
\ref{Figure9}(a) and \ref{Figure9}(b). (Note that the peak appearing
in the polarization at $\varepsilon _d =0$ is due to a van Hove
singularity in density of states for a two dimensional square
lattice in the tight binding Hamiltonian.)

Finally, we note that recent experiments \cite{Hill_2006} on a
graphene-based spin valve did lead to a significant $MR$  for
undoped graphene.  Our results show that this result cannot be
understood purely on the basis of a clean, non-interacting
electron model. However, the ferromagnetic contacts reported by
Hill {\it et al.} in Ref. \onlinecite{Hill_2006} exhibited
strongly non-Ohmic, nonlinear I-V behavior which one would not
usually associate with normal metallic contacts
\cite{geim_private}. When similar devices with Ohmic contacts
(using a Ti sublayer) were fabricated, the magnetoresistance
previously observed completely disappeared \cite{geim_private}.
These experimental findings support our results that spin valves
employing clean graphene and simple, homogeneously magnetized
leads should have relatively very low $MR$.

\section{Summary and Conclusion}
In this work we undertook a detailed study of the conduction
properties of wide graphene strips, with two different models for
the source and drain leads.  We reconfirmed that for undoped
graphene, the system can be described by a {\it conductivity} in
the $L \rightarrow \infty$ limit even when defects are absent from
the system, and examined this behavior with respect to a broad
range of lead parameters.  The resulting conductance turns out to
be relatively insensitive to these.

We then used these results to compute the magnetoresistance of a
simple three stripe  spin-valve device with graphene acting as the
non-magnetic material between the ferromagnetic leads.  Two types of
ferromagnetic lead systems were considered: one with a single ($s$)
orbital for each spin state, with band centers separated in energy
to induce spin polarization, and another with a narrow $d$ band
which was taken to be spin-polarized. It was found that the
magnetoresistance $MR$ was rather small for most circumstances in
both cases, largely due to the insensitivity of the conductivity
with respect to conditions in the leads.

{\bf Acknowledgements.} The authors thank  J.J.Palacios,
J.Fern\'andez-Rosier, T.Stauber, B.Wunsch, E.Prada and P.San-Jos\'e
for useful discussions.
The authors are also grateful to Andr\'e Geim from discussions about
the results of Ref. \onlinecite{Hill_2006}.
This work was supported by MAT2006-03741
(Spain) (LB), by the NSF through Grant Nos. DMR-0454699 and
DMR-0704033 (HAF).


\end{document}